\documentclass[pra,twocolumn]{revtex4-1}
\usepackage{multirow}
\usepackage{amsmath,amssymb,bbold,bm,soul}
\usepackage[pdftex]{graphicx}
\usepackage{epstopdf}
\usepackage{color}

\newcommand{\beq}{\begin{equation}}
\newcommand{\eeq}{\end{equation}}

\newcommand{\by}[1]{#1 \times #1}
\newcommand{\rk}{{\rm Rank}}
\newcommand{\tr}{{\rm Tr}}

\newcommand{\comment}[1]{}

\begin{document}
\title{The power of being positive: Robust state estimation made possible by quantum mechanics}

\author{Amir Kalev}
\email{amirk@unm.edu}
\author{Charles H. Baldwin}
\email{baldwin4@unm.edu}
\author{Ivan H. Deutsch}
\email{ideutsch@unm.edu}
\affiliation{Center for Quantum Information and Control, MSC07--4220, University of New Mexico, Albuquerque, New Mexico 87131-0001, USA}
\date{\today}

\begin{abstract}
We study the problem of quantum-state tomography under the assumption that the state of the system is close to pure. In this context, an efficient measurements that one typically formulates uniquely identify a pure state from within the set of other pure states. In general such measurements are not robust in the presence of measurement noise and other imperfections, and therefore are less practical for tomography. We argue here that state tomography experiments should instead be done using measurements that can distinguish a pure state from {\em any} other quantum state, of any rank. We show that such nontrivial measurements follows from the physical constraint that the density matrix is positive semidefinite and prove that these measurements yield a robust estimation of the state. We assert that one can implement such tomography relatively simply by measuring only a few random orthonormal bases; our conjecture is supported by numerical evidence. These results are generalized for estimation of states close to bounded-rank.
\end{abstract}

\maketitle

\section{Introduction} 
Quantum-state tomography (QST) is the standard protocol used to estimate and characterize the state of a quantum system. QST is expensive to implement experimentally, however, since the resources to reconstruct an arbitrary state scale poorly with the dimension of the Hilbert space.  Nevertheless, one can improve the efficiency of QST through the use of prior information.  In state-of-the-art experiments in quantum information science the goal is not to produce arbitrary states but pure states. If one has calibrated the device and seen it to be performing well, e.g., based on randomized benchmarking~\cite{emerson05,knill08,magesan11}, then with good confidence we can expect that the device is processing states that are close to pure.  As a first approximation, we can assume the state is exactly pure, though of course, this is never true in any real device. Including this prior information in QST results in more manageable tomography protocol where resources scale only linearly with the dimension as has been studied in a variety of previous works~\cite{Flammia05, Finkelstein04, Heinosaari13, Chen13, Carmeli14, Goyeneche14, Carmeli15, Kalev15,Kech15, Kech15b, Baldwin15}. 

When considering pure-state tomography, and bounded-rank state (a state with rank $\leq r$) tomography more generally, a natural notion of informational completeness emerges~\cite{Heinosaari13}, referred to as {\em rank-$r$ completeness}. A measurement is rank-$r$ complete if the measurement probabilities uniquely distinguish a state with rank $\leq r$ from any other state with rank $\leq r$. A rigorous definition is given below.  Recently, there was an effort to formulate such measurements~\cite{Flammia05,Finkelstein04,Heinosaari13,Carmeli15,Goyeneche14,Kech15}. For example, Heinosaari~{\em et al.}~\cite{Heinosaari13} showed for states in finite $d$-dimensional Hilbert space, the expectation values of particular $4r(d-r)$ observables corresponds to rank-$r$ complete measurement. The set of quantum states with rank $\leq r$ is, however, nonconvex and therefore, in the presence of experimental noise, convergence to a reliable estimate is not guaranteed when using rank-$r$ complete measurements. This poses a concern for the practicality of such measurements for QST.

The purpose of this contribution is two fold: (i) We develop the significance of a different notion of informational completeness, {\em rank-$r$ strict-completeness}. We prove that strictly-complete measurements allow for robust estimation of bounded-rank states in the realistic case of noise and experimental imperfections, by solving essentially any convex program. Because of this, strictly-complete measurements are crucial for the implementation of pure-state QST.  (ii) We argue, based on numerical evidence, that it is fairly straightforward to implement such measurements based solely on projective measurements in few random orthonormal bases,with a very weak dependence on the dimension of Hilbert space. 

The remainder of this article is organized as follows: In Sec.~\ref{sec:Info Complete}, we define the notions of rank-$r$ complete and rank-$r$ strictly-complete POVMs. In Sec.~\ref{sec:Power of strict}, we establish the importance of strictly-complete POVMs for practical QST. In Sec.~\ref{sec:numerics}, we provide numerical evidence that a measurement of a few orthonormal bases implements a strictly-complete POVM  and demonstrate the robustness of such measurements. Finally in Sec.~\ref{sec:conclusions}, we offer some conclusions.

\section{Informational completeness in bounded-rank QST} \label{sec:Info Complete}
\begin{figure}[t]
\centering
\includegraphics[width=\linewidth]{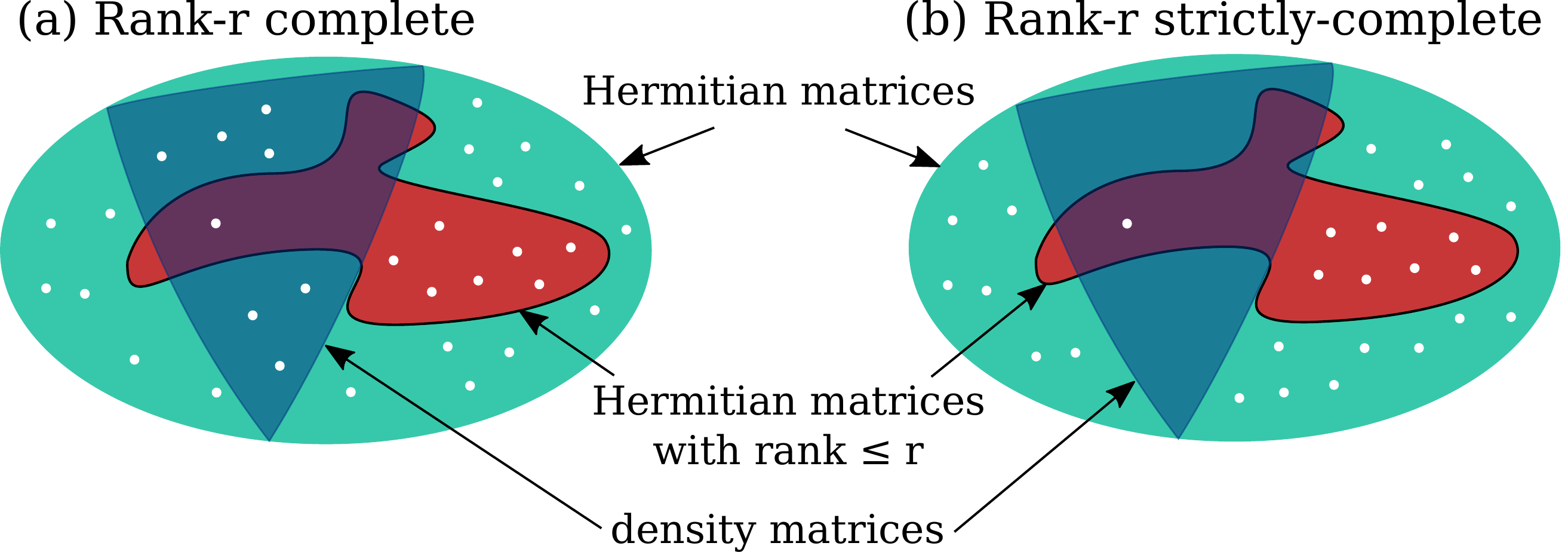}
\caption{{\bf Various notions of completeness in bounded-rank QST.} The white dots represent Hermitian matrices, physical or not, that are consistent with the (noiseless) measurement record.  {\bf (a) Rank-$r$ completeness.} The measurement record, distinguishes the rank $\leq r$ state from any other rank $\leq r$ PSD matrix. However, there generally will be infinitely many other states,  with rank greater than $r$, that are consistent with the measurement record.  {\bf (b)  Rank-$r$ strict-completeness.}  The measurement record  distinguishes the rank $\leq r$ state from any other PSD matrix. Thus it is unique in the convex set of positive matrices.  Once normalized, these are possible quantum states.  Thus, the only one physical quantum state is consistent with the measurement record.}
\label{fig:illustration}
\end{figure}
QST has two basic ingredients, states and measurements, so it is important to define these precisely.  A quantum state is a density matrix, $\rho$, that is positive semidefinite (PSD) and normalized to unit trace. A quantum measurement with $m$ possible outcomes (events) is defined by a POVM with $m$ elements,  ${\cal E}=\{E_\mu:E_\mu \geq 0,\, \sum_{\mu=1}^m E_\mu=\mathbb{1}\}$. A POVM is informationally complete if the measurement probabilities, $\tr(\rho E_\mu)$, distinguish the state $\rho$ from any other state. While we typically think of POVMs as as a map between quantum states and probabilities, mathematically,  a POVM is a map on any Hermitian matrix, and in particular on PSD matrices. In fact, it is advantageous  to define the different notions of informational completeness, rank-$r$ complete and rank-$r$ strictly-complete, in terms of PSD matrices, and then apply them to quantum states, i.e., PSD matrices with unit trace. This highlights the fact that the definitions and our results are independent of the trace constraint of quantum states, and only depend on their positivity property. Additionally, because, by definition, the POVM elements sum to the identity, the trace of the matrix is always ``measured.''  Therefore, we first present the mathematical formalism in terms of PSD matrices, and then apply it to quantum states. We will comment about the role of the trace constraint where necessary.

Consider a $d$-dimensional Hilbert space, ${\mathcal H}_d$. Let ${\cal M}_{\cal E}[\cdot]=(\tr(\cdot E_1),\ldots,\tr(\cdot E_m))$ be the map defined by the POVM from Hermitian matrices to a real vector space, ${\mathbb R}^m$. When ${\cal M}_{\cal E}$ acts on a quantum state $\rho$, it returns a vector of probabilities, ${\cal M}_{\cal E}[\rho]=(\tr(\rho E_1),\ldots,\tr(\rho E_m))\equiv{\bm p}$. More generally, when ${\cal M}_{\cal E}$ acts on a PSD matrix $X$, it returns a ``measurement'' vector ${\bm y}$, whose components $y_\mu\geq0$, and $\sum_{\mu=1}^m y_\mu=\tr X$. It is also useful to define the kernel of a POVM, ${\rm Ker}({\mathcal E})\equiv \{X:{\cal M}_{\cal E}[X]={\bf 0}\}$. Since the POVM elements sum to the identity matrix, we immediately obtain that every $X\in{\rm Ker}({\mathcal E})$ is traceless, $\tr(X)=0$.

In bounded-rank QST, rank-$r$ completeness is a natural concept~\cite{Heinosaari13, Carmeli14, Kalev15}, which we define here in terms of PSD matrices.\vspace{0.2cm}\\
\noindent {\bf Definition~1:} Let  ${\cal S}_r=\{X:X\geq0,\rk X\leq r\}$ be the set of $\by{d}$ PSD matrices with rank $\leq r$.  A POVM is said to be {\em rank-$r$ complete} if  
\begin{equation} 
\label{restricted_definition}
\forall \, X_1, X_2 \in {\cal S}_r, X_1\neq X_2, \, {\cal M}_{\cal E}[X_1]\neq{\cal M}_{\cal E}[X_2],
\end{equation}
except for possibly a set of rank-$r$ states that are dense on a set of measure zero, called the ``failure set.''
\vspace{0.2cm}\\
In the context of QST, the probabilities of a rank-$r$ complete POVM uniquely identify the rank $\leq r$ state from within the set of all PSD matrices with rank $\leq r$, ${\cal S}_r$. Figure~\ref{fig:illustration}a illustrates the notion of rank-$r$ completeness. Given a rank $\leq r$ state in  the failure set, the probabilities of the measurement outcomes do not uniquely identify the state from within the set of all rank $\leq r$ states; this was considered in~\cite{Flammia05,Goyeneche14,Baldwin15}.  For random states or a random measurement bases, however, the chance of hitting the failure state is vanishingly small. We comment on the implications of the failure set for practical tomography in the sections to follow.

Using ${\rm Ker}({\mathcal E})$ we arrive at an alternative, equivalent, definition for rank-$r$ complete:  A POVM ${\mathcal E}$ is rank-$r$ complete if $\forall \, X_1,X_2 \in {\cal S}_r$, with $\, X_1\neq X_2$, the difference $\Delta= X_1-X_2$ is not in the kernel of ${\mathcal E}$, i.e., $\exists \, E_\mu$ such that $\tr(E_\mu \Delta)\neq0$. Carmeli~{\em et al.}~\cite{Carmeli14} showed that a necessary and sufficient condition for a POVM to be rank-$r$ complete is that every nonzero $\Delta\in{\rm Ker}({\mathcal E})$ has $\max(n_{-},n_{+})\geq r+1$, where $n_{+}$ and $n_{-}$ are the number of strictly positive and strictly negative eigenvalues of a matrix, respectively. This condition was derived for PSD matrices with rank $\leq r$. If we exclude the positivity property, and only consider the rank property, we obtain a sufficient condition:  a POVM is rank-$r$ complete if every nonzero $\Delta\in{\rm Ker}({\mathcal E})$ has $\rk(\Delta)\geq 2r+1$. This sufficient condition applies to all Hermitian matrices with rank $\leq r$.  Using the sufficient condition alone, it was shown~\cite{Heinosaari13} that the expectation values of particular $4r(d-r)$ observables corresponds to rank-$r$ complete measurement, and moreover~\cite{Kech15} that a measurement of $4r\lceil\frac{d-r}{d-1}\rceil$ random orthonormal bases is rank-$r$ complete. 

Definition~1 does not say anything about higher-rank matrices. In particular for QST, if the POVM is rank-$r$ complete, and the state of the system has rank $\leq r$, then according to the definition, there could be higher-rank quantum states that yeild the same probabilities. Therefore, when applying a rank-$r$ complete measurement, in the presence of noise and other experimental imperfections, it may be difficult to differentiate between the unique rank $\leq r$ states and these higher-rank states since the set ${\cal S}_r$ is nonconvex. To overcome this difficulty, we can consider a ``stricter" definition of rank-$r$ completeness~\cite{Chen13,Carmeli14, Kalev15}\vspace{0.2cm}:\\
\noindent {\bf Definition~2:} Let  ${\cal S}=\{X:X\geq0\}$ be the set of $\by{d}$ PSD matrices. A POVM is said to be {\em rank-$r$ strictly-complete} if
\begin{equation} 
\label{strictly_definition}
 \forall \, X_1 \in {\cal S}_r, \, \forall \, X_2 \in {\cal S}, X_1\neq X_2, \, {\cal M}_{\cal E}[X_1]\neq{\cal M}_{\cal E}[X_2],
\end{equation}
except for possibly a set of rank-$r$ states that are dense on a set of measure zero, called the ``failure set.''
\vspace{0.2cm}\\
The implication for QST is that when the state being measured is promised to be in ${\cal S}_r$, the measurement probabilities of a rank-$r$ strictly-complete POVM distinguish this state from any other PSD matrix, of {\em any} rank (except on the failure set). Figure~\ref{fig:illustration}(b) illustrates the notion of rank-$r$ strict-completeness. 

Carmeli~{\em et al.}~\cite{Carmeli14} showed that a POVM is rank-$r$ strictly-complete if, and only if, every nonzero $X\in{\rm Ker}({\mathcal E})$ has $\min(n_{-},n_{+})\geq r+1$. Again, this condition relies on the PSD property of the matrices. To date, there are only a few known POVMs that are rank-$r$ strictly-complete. One example, we recently showed that a specific set of $4r+1$ orthonormal bases is rank-$r$ strictly-complete measurement in the context of matrix completion~\cite{Baldwin15}.  

In contrast to the notion of rank-$r$ completeness, which can be defined generally for bounded-rank Hermitian matrices, the definition of strict-completeness applies nontrivially only to PSD matrices, and therefore, in particular to quantum states~\cite{Baldwin15}. To see this, let us apply the definition of strict-completeness for bounded-rank Hermitian matrices, ignoring positivity. Let $R$ be a Hermitian matrix with $\rk(R)\leq r$ and let $H$ be a Hermitian matrix with $\rk(H)\leq d$. Then the rank of their difference, $\Delta=R-H$, is at most $d$, $\rk(\Delta)\leq d$. Therefore, following the definition, in order to be a rank-$r$ strictly-complete for Hermitian matrices, the kernel of the measurement must not include Hermitian matrices whose rank $\leq d$. Since the latter is the set of all $d\times d$ Hermitian matrices, we conclude that the kernel of such measurement includes only the (trivial) zero matrix. This, in turn, means that to be a rank-$r$ strictly-complete measurement for all Hermitian matrices the measurement must be fully informationally complete, i.e., able to uniquely identify any Hermitian matrix, of any rank.

\section{The power of strictly-complete measurements} ~\label{sec:Power of strict}
The power of strictly-complete measurements become evident when we consider implementations in  a realistic experimental context. It is essential that the estimation protocol be robust to noise and other imperfections. Thus, any realistic estimation procedure should allow one to find the closest estimate (by some appropriate measure of distance) given a bound on the noise. We can address this by convex optimization, whereby the estimate is found by minimizing a convex function over a convex set. Since rank-$r$ strictly-complete POVMs uniquely identify a rank-$r$ quantum state within the {\em convex} set of PSD matrices, the data obtained from measurements defined by these POVMs fits the convex optimization paradigm. On the other hand, rank-$r$ complete measurements uniquely identify the rank-$r$ state only within the {\em non-convex} set of rank-$r$ PSD matrices, and therefore are not compatible to use with convex optimization. 

In the following corollaries the variable $X$ is a PSD matrix, not necessarily with a unit trace. We first consider the case that the measurement is noiseless with the following corollary:
\vspace{0.1cm}\\
\noindent {\bf Corollary~1 (uniqueness):} Let $\rho_0$ be a quantum state with rank $\leq r$, and let ${\bm p}= {\cal M}_{\cal E}[\rho_0]$ be the corresponding probabilities of a rank-$r$ strictly-complete POVM ${\cal E}$. Then, the solution to
\begin{equation}
\label{general_positive_CS}
\hat{X} = \arg\min_X \mathcal{C}(X)\;\; {\rm s.t.}\; {\cal M}[X]=\bm{p}\, \,  {\rm and} \, \,  X \geq 0,
\end{equation}
or to
\begin{equation}
\label{general_norm_positive_CS}
\hat{X} = \arg\min_X \Vert{\cal M}[X]-\bm{p}\Vert\;\; {\rm s.t.}\; X \geq 0,
\end{equation}
where $\mathcal{C}(X)$ is a any convex function of $X$, and $\Vert\cdot\Vert$ is any norm function, is unique:  $\hat{X} = \rho_0$.\vspace{0.1cm}\\
\noindent {\em Proof:} This is a direct corollary of the definition of strict-completeness. Since, by definition, the probabilities of rank-$r$ strictly-complete POVM uniquely determine $\rho_0$ from within the set of all PSD matrices, its reconstruction becomes a feasibility problem over the convex set $\{{\cal M}[X]=\bm{p},X\geq0\}$,
\begin{equation} \label{feasibility}
{\rm find}\; X\;\; {\rm s.t.}\; {\cal M}[X]=\bm{p}\, \,  {\rm and} \, \, X \geq 0.
\end{equation}
The solution for this feasibility problem is $\rho_0$ uniquely. Therefore, any convex program, which looks for the solution within the feasible set, is guaranteed to find $\rho_0$.\hfill $\square$

Corollary~1 was proved in~\cite{Kech15b} for the particular choice $\mathcal{C}(X)=\tr(X)$ , and also in the context of compressed sensing measurements in~\cite{Kalev15}. Note, while one can also include a trace constraint $\tr(X) = t$, in this noiseless case, Eqs.~\eqref{general_positive_CS} and~\eqref{general_norm_positive_CS}, this is redundant since any POVM measures the trace of a matrix.  Thus, if we have prior information that $\tr(X)=t$, then the feasible set in Eq.~\eqref{feasibility} is equal to the set $\{ X \, | \,{\cal M}[X]=\bm{p}, \, X \geq 0,\,  {\rm and} \, \tr(X) = t \}$.  In particular, in the context of QST with noiseless data, the constraint $\tr\rho=1$ would be redundant; the reconstructed state would necessarily be properly normalized.

The Corollary implies that strictly-complete POVMs allow for the reconstruction of bounded-rank states via convex optimization, even though the set of bounded-rank states is nonconvex. Moreover, all convex programs over the feasible solution set, i.e., of the form of Eqs.~\eqref{general_positive_CS} and~\eqref{general_norm_positive_CS}, are equivalent for this task. For example, this result applies to maximum-(log)likelihood estimation~\cite{Hradil97} where $\mathcal{C}(\rho) =-\log( \prod_{\mu}\tr(E_\mu\rho)^{p_\mu})$.  The Corollary does not apply for states in the POVM's failure set, if such set exists. 

In any real experiment, the assumption that the state has rank $\leq r$ is only an approximation, and moreover the measurement record always contains experimental noise. In this case, we wish to find an estimate of the state, e.g., the closest state consistent the measurement record. Determining such a state for rank-$r$ complete measurements is generally a hard problem since the set of bounded-rank states in nonconvex.  However, strict-completeness, together with the convergence properties of convex programs, ensure a robust state estimation in realistic experimental scenarios. This is the main advantage of strictly-complete measurements and it is formalized in the following corollary.  \vspace{0.1cm}\\
\noindent {\bf Corollary~2 (robustness):} Let $\sigma$ be the state of the system, and let ${\bm f}= {\cal M}_{\cal E}[\sigma]+{\bm e}$ be the (noisy) measurement record of a rank-$r$ strictly-complete POVM ${\cal E}$, such that $\Vert{\bm e}\Vert\leq\epsilon$. If  $\Vert{\bm f}-{\cal M}_{\cal E}[\rho_0]\Vert\leq \epsilon$ for some quantum state $\rho_0$ with $\rk(\rho_0)\leq r$, then the solution to
\begin{equation}
\label{general_positive_CS_noisy}
\hat{X} = \arg\min_X \mathcal{C}(X)\;\; {\rm s.t.}\; \Vert{\cal M}[X]-\bm{f}\Vert\leq\epsilon\, \,  {\rm and} \, \,  X \geq 0,
\end{equation}
or to
\begin{equation}
\label{general_norm_positive_CS_noisy}
\hat{X} = \arg\min_X \Vert{\cal M}[X]-\bm{f}\Vert\;\; {\rm s.t.}\; X \geq 0,
\end{equation}
where $\mathcal{C}(X)$ is a any convex function of $X$, is robust:  $\Vert\hat{X} - \rho_0\Vert\leq C_{\cal E}\epsilon$, and $\Vert\hat{X} - \sigma\Vert\leq 2C_{\cal E}\epsilon$, where $\Vert\cdot\Vert$ is any $p$-norm, and $C_{\cal E}$ is a constant which depends only on the POVM.\vspace{0.1cm}\\
\\
\noindent The proof, given in the Appendix, is derived from Lemma~V.5 of~\cite{Kech15b}  where it was proved for the  particular choice $\mathcal{C}(X)=\tr(X)$. In~\cite{Kalev15} this was also studied in the context of compressed sensing measurements. This corollary assures that if the state of the system is close to a bounded-rank density matrix, and it was measured with strictly-complete measurements, then it can be robustly estimated with any convex program, constrained to the set of PSD matrices. In particular, it implies that all convex estimators preform qualitatively the same for low-rank state estimation. This may be advantageous especially when considering estimation of high-dimensional quantum state. As in the noiseless case, the trace constraint is not necessary for Corollary~2, and in fact leaving it allows us to make different choices for $\mathcal{C}(X)$, as was done in Ref.~\cite{Kech15b}. Therefore, in general, the estimated matrix $\hat{X}$ is not normalized, $\tr\hat{X}\neq1$. The final estimation of the state is then given by $\hat{\rho} = \hat{X}/\tr(\hat{X})$. In principle, we can consider a different version of Eqs.~\eqref{general_positive_CS_noisy} and~\eqref{general_norm_positive_CS_noisy} where we include the trace constraint, and this may have implications for the issue of ``bias" in the estimator. This will be studied in more details elsewhere.

\section{Numerical experiments} \label{sec:numerics}
We study here a straightforward experimentally feasible protocol to implement strictly-complete measurements. In particular, in our numerical analysis shown below, we find that simply by measuring only few random orthonormal bases amounts to strict-completeness. Note, measurement of random bases were also considered in the context of compressed sensing (see, e.g., in~\cite{Kueng15,Acharya15}). However, when taking into account the PSD property of density matrices, we obtain strict-completeness with fewer measurements than required for compressed sensing~\cite{Kalev15}. Therefore, strict-completeness does not imply compressed sensing. While for quantum states, all compressed sensing measurements are strictly-complete~\cite{Kalev15}, not all strictly-complete measurements are compressed sensing measurements.

As a first numerical test we study the number of random bases that corresponds to a rank-$r$ strictly-complete measurement for $r=1,2,3$.  To determine this, we take the ideal case where the probabilities are known exactly and the rank of the state is fixed. We consider two types of measurements on a variety of different dimensions: (i) a set of  Haar-random orthonormal bases on unary systems with dimensions $d=11, 16, 21, 31, 41$, and $51$; and (ii) a set of local Haar-random orthonormal bases on a tensor product of $n$ qubits with $n=3,4,5$, and $6$, corresponding to $d=8, 16, 32$, and $64$, respectively. For each dimension, and for each rank, we generate $25d$ Haar-random states. Each state is then measured with an increasing number of bases. The noiseless measurement record, $\bm{p}$, are the probabilities of the measurement outcomes. After each new basis measurement we use the constrained least-square (LS) program, Eq.~\eqref{general_norm_positive_CS} where $\Vert\cdot\Vert$ is the $\ell_2$-norm, to produce an estimate of the state. We emphasize that the constrained LS finds the quantum state that is the most consistent with $\bm{p}$ with no restrictions on rank. The procedure is repeated until all estimates match the states used to generate the data (up to numerical error of $10^{-5}$ in infidelity). This indicates the random bases used correspond to a rank-$r$ strictly-complete POVM.

\begin{table}[h]
\centering
\begin{tabular}{ cc|c|c|c|c|c|||c|c|c|c| }
& \multicolumn{9}{c}{\bf{Dimension}}\\ \cline{2-11}
 & \multicolumn{6}{|c|||}{Unary}  & \multicolumn{4}{c| }{Qubits}\\
 \multicolumn{1}{ c|| }{\bf{Rank}} &\bf{11} &\bf{16}&\bf{21} &\bf{31} &\bf{41}&\bf{51} &\bf{8} &\bf{16} &\bf{32}&\bf{64} \\
\hline\hline
\multicolumn{1}{ |c|| }{\bf{1}} &\multicolumn{6}{c|||}{6} & \multicolumn{4}{c|}{6}  \\\cline{1-11} 
\multicolumn{1}{ |c|| }{\bf{2}} & 7  & 8  & 8  & \multicolumn{3}{ c||| } {9}& \multicolumn{2}{ c| } {9} &  \multicolumn{2}{ c| } {10}  \\\cline{1-11}
 \multicolumn{1}{ |c|| }{\bf{3}} & 9  & 10  & 11  & 12 & 12  & 13 & 12 & \multicolumn{3}{ c| } {15}\\\cline{1-11}
\hline
\end{tabular}
\caption{{\bf Number of random orthonormal bases corresponding to strict-completeness.}  Each cell lists the minimal number of measured bases for which the infidelity was below $10^{-5}$  for each of the tested states in the given dimensions and ranks. This indicates that a measurement of only few random bases is strictly-complete POVM.}\label{tbl:noiseless}
\end{table}  

We present our findings in Table~\ref{tbl:noiseless}.  For each dimension, we also tested fewer bases than listed in the table. These bases corresponded to strictly-complete measurements for most states but not all. For example, for the unary system with $d=21$, using the measurement record from $5$ bases we can reconstruct all states with an infidelity below the  threshold except for one. The results indicate that measuring only few random bases, with weak dependence on the dimension, corresponds to a strictly-complete POVM for low-rank quantum states. Moreover, the difference between, say rank-1 and rank-2, amounts to measuring only few more bases. This is important, as discussed below, in realistic scenarios when the state of the system is known to be close to pure. Finally, when considering local measurements, more bases are required to account for strict-completeness when compared to unary system, see for example results for $d=16$.  

Next, we simulate a realistic scenario, where the state of the system has high purity and the experimental data contains statistical noise. From Corollary~2 we expect to obtain a robust estimation of the state, by solving any convex estimator of the form of Eqs.~\eqref{general_positive_CS_noisy} and~\eqref{general_norm_positive_CS_noisy}, when measuring a set of random orthonormal bases that correspond to rank-$1$ strictly-complete POVM. 

As before we numerically study unary systems, ($d=11, 21$, and $31$) and systems composed of (3,4, and 5) qubits, with the same bases introduced for the noiseless case. We generate 200 Haar-random pure-states (target states), $\{| \psi\rangle\}$, and take the state of the system to be $\sigma =(1- q) | \psi \rangle \langle \psi | + q \tau$, where $q = 10^{-3}$, and $\tau$ is a random full-rank state. The measurement record, $\bm{f}$, is simulated by sampling $m = 300 d$ trials from the probability distribution corresponding to the measured basis. For each number of measured bases, we estimate the state~\cite{cvx,Hradil97} based on different convex optimization programs: (i) A constrained trace-minimization program, Eq.~\eqref{general_positive_CS_noisy} with $\mathcal{C}(\rho)=\tr(\rho)$,  (ii) a constrained LS program, Eq.~\eqref{general_norm_positive_CS_noisy} using the $\ell_2$-norm, and  (iii) the maximum-(log)likelihood program based on the algorithm described in~\cite{teo11}.

\begin{figure}[t]
\centering
\includegraphics[width=\linewidth]{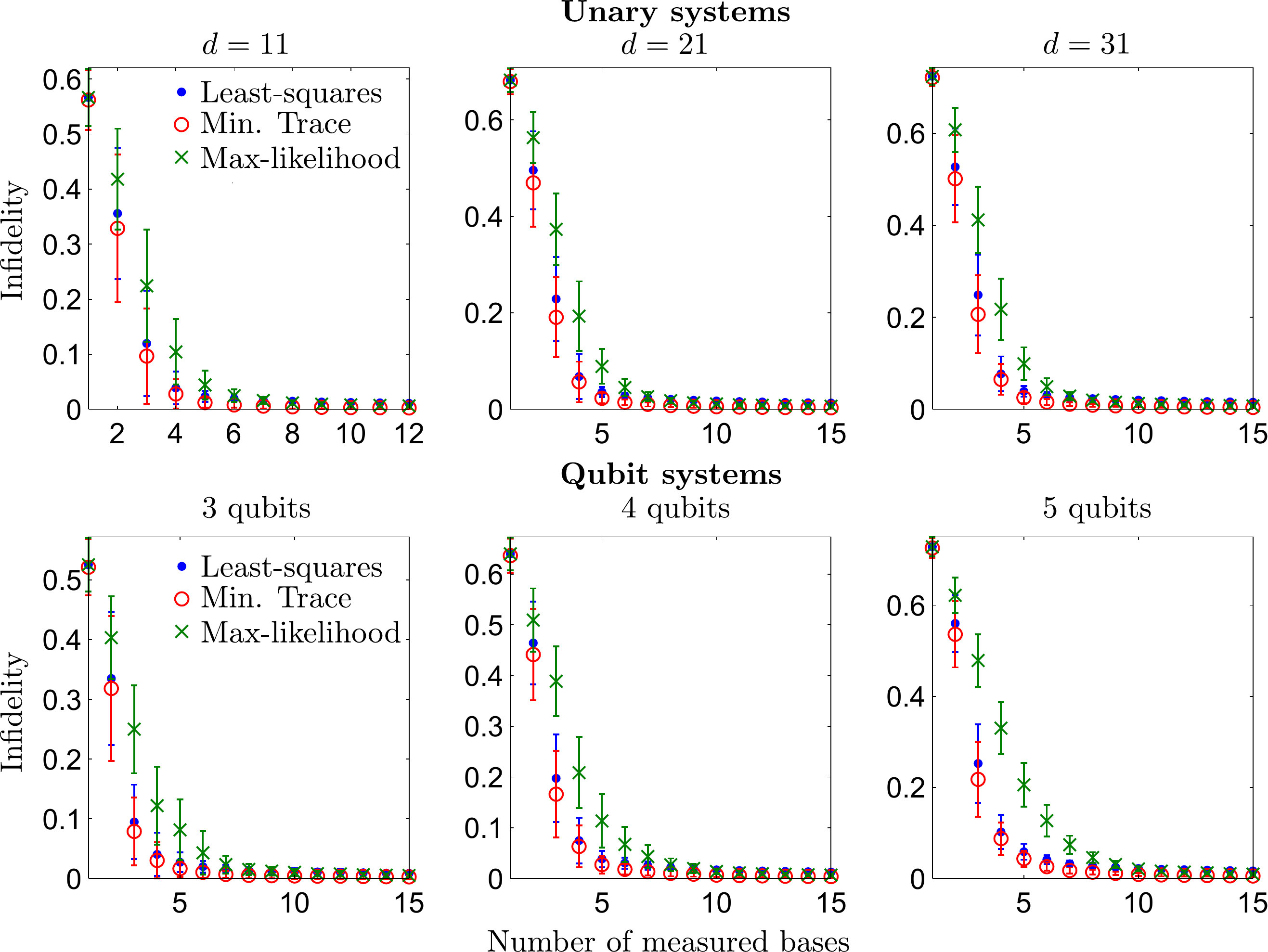}
\caption{{\bf Simulation of QST under realistic conditions.} We assume that the state of the system is close to a target pure state. We plot the average infidelity between the target pure state and its estimation as a function of measured bases. Most of the information about the state is obtained when the number of measured bases corresponds to rank-1 strictly-complete, {\em c.f.} Table~\ref{tbl:noiseless}. Following Corollary~2, we obtain a robust estimation regardless of the particular program used to estimate the state.}
\label{fig:noisy}
\end{figure}

In Fig.~\ref{fig:noisy} we plot the average infidelity (over all tested states) between the target state, $ | \psi \rangle$, and its estimation, $\hat\rho$, $1-\overline{\langle\psi|\hat\rho| \psi \rangle}$. As ensured by Corollary~2, the various convex programs we used robustly estimate the state with number of bases correspond to rank-1 strictly-complete POVM.  Furthermore, in accordance to our findings of Table~\ref{tbl:noiseless}, measuring only a few more bases, such that the overall POVM is rank-$2$ strictly-complete, and so forth, we improve the estimation accordingly. A practical protocol for doing QST could be to iteratively measure a basis and reconstruct the state via convex optimization until the fidelity converges. 

\section{Conclusions} \label{sec:conclusions}
We have studied QST under the assumption that the state of the system is known to be close to a pure state, or more generally, to a rank $\leq r$ state. Since the set of rank $\leq r$ states is nonconvex, it is generally difficult to robustly estimate the state of the system by measuring rank-$r$ complete POVMs. We showed, however, that a robust estimation is guaranteed if the measurements are rank-$r$ strictly-complete. Such measurements  efficiently identify a low-rank state from within the set of all quantum states. The essential ingredient to strict-completeness is the positivity constraint associated with physical density matrices. Moreover, the estimation can be done by solving any convex program over the feasible set and the estimate returned is robust to errors. In addition, guided by numerical tests, we conjecture that it is rather straightforward to design strictly-complete measurements. Depending on the dimension and rank of the state, measuring only few random orthogonal bases results in strict-completeness, for unary system as well as system of qubits. Therefore, strictly-complete measurements are essential for practical implementations of QST.

\acknowledgements
This work was supported by NSF Grants PHY-1212445, PHY-1521016, and PHY-1521431.

\appendix
\section{Proof of Corollary~2}
The proof of Corollary~2 uses Lemma~V.5 of~\cite{Kech15b}, restated as follows.\\
\noindent {\bf Lemma~3:} Let ${\cal E}$ be a rank-$r$ strictly-complete POVM, and let ${\bm f}= {\cal M}_{\cal E}[\sigma]+{\bm e}$ be the measurement record of some quantum state, $\sigma$. If  $\Vert{\bm f}-{\cal M}_{\cal E}[\rho_0]\Vert\leq \epsilon$ for some quantum  state $\rho_0$ with $\rk\rho_0\leq r$,   then for every PSD matrix $X$ such that $\Vert{\cal M}_{\cal E}[X]-{\bm f}\Vert\leq \epsilon$, we have $\Vert X-\rho_0\Vert\leq C_{\cal E}\epsilon$, where $C_{\cal E}$ depends only on the POVM.\\

The proof of this Lemma can be found in~\cite{Kech15b}. To prove Corollary~2, we first show that $\Vert\hat{X}-\rho_0\Vert\leq C_{\cal E}\epsilon$. The convex programs of Eqs.~\eqref{general_positive_CS_noisy} and~\eqref{general_norm_positive_CS_noisy} in the main text look for a solution that minimizes some convex function on the set  $\{\Vert{\cal M}_{\cal E}[X]-{\bm f}\Vert\leq \epsilon, X\geq0\}$. According to the Lemma,  any PSD matrix $X$ within this set satisfies $\Vert X-\rho_0\Vert\leq C_{\cal E}\epsilon$. Since the  solution $\hat{X}$ is also in that set, we obtain that $\Vert\hat{X}-\rho_0\Vert\leq C_{\cal E}\epsilon$. 

Next, we show that $\Vert\hat{X}-\sigma\Vert\leq 2C_{\cal E}\epsilon$. Since  we assume $\Vert{\bm e}\Vert\leq\epsilon$, $\sigma$ is in the set  $\{\Vert{\cal M}_{\cal E}[X]-{\bm f}\Vert\leq \epsilon, X\geq0\}$, and according to the Lemma $\Vert\sigma-\rho_0\Vert\leq C_{\cal E}\epsilon$. Therefore we have
\begin{align*}
\Vert\hat{X}-\sigma\Vert&=\Vert\hat{X}-\rho_0-\sigma+\rho_0\Vert\leq\Vert\hat{X}-\rho_0\Vert+\Vert\sigma-\rho_0\Vert\\&\leq 2C_{\cal E}\epsilon.
\end{align*}
\hfill$\square$\\
For convenience one parameter, $\epsilon$, is used to quantify the various bounds. However it is straightforward to generalize this result to the case where the bounds are quantified by different values.


\begin{thebibliography}{99}
\bibitem{emerson05} J. Emerson, R. Alicki, and K. \.{Z}yczkowski, J. Opt. B {\bf 7}, S347 (2005).
\bibitem{knill08} E. Knill, D. Leibfried, R. Reichle, J. Britton, R. B. Blakestad, J. D. Jost, C. Langer, R. Ozeri, S. Seidelin, and D. J. Wineland,  Phys. Rev. A {\bf 77}, 012307 (2008).
\bibitem{magesan11} E. Magesan, J. M. Gambetta, and J. Emerson,  Phys. Rev. Lett.  {\bf 106}, 180504 (2011).
%Randomized benchmarking of quantum gates
\bibitem{Flammia05} S. T. Flammia, A. Silberfarb, and C M. Caves,  Found. Phy., {\bf 35}, 1985 (2005).
%Minimal Informationally Complete Measurements for Pure States
\bibitem{Finkelstein04} J. Finkelstein, Phys. Rev. A {\bf 70}, 052107 (2004).
%Pure-state informationally complete and “really” complete measurements
\bibitem{Heinosaari13} T. Heinosaari, L. Mazzarella, and M. M. Wolf, Commun. Math. Phys. {\bf 318}, 355-374 (2013).
%Quantum Tomography under Prior Information
\bibitem{Chen13} J. Chen, H. Dawkins, Z. Ji, N. Johnston, D. Kribs, F. Shultz, and B. Zeng, Phys. Rev. A {\bf 88}, 012109 (2013).
%Uniqueness of Quantum States Compatible with Given Measurement Results
\bibitem{Carmeli14} C. Carmeli, T. Heinosaari, J. Schultz, and A. Toigo, J. Phys. A: Math. Theor. {\bf 47}, 075302 (2014).
%Tasks and premises in quantum state determination
\bibitem{Goyeneche14} D. Goyeneche, G. Ca{\~n}as, S. Etcheverry, E. S. G{\'o}mez, G. B. Xavier, G. Lima, and A. Delgado, Preprint arXiv:1411.2789 (2014).
%Five measurement bases determine pure quantum states on any dimension
\bibitem{Carmeli15} C. Carmeli, T. Heinosaari, J. Schultz, and A. Toigo, Preprint  arXiv:1504.01590 (2015).
%How many orthonormal bases are needed to distinguish all pure quantum states?
\bibitem{Kalev15}A. Kalev, R. L. Kosut, and I. H. Deutsch,  {\em npj} Quantum Information {\bf 1}, 15018 (2015).
%Quantum Tomography Protocols with Positivity are Compressed Sensing Protocols
\bibitem{Kech15} M. Kech and M. M. Wolf, Preprint  arXiv:1507.00903 (2015).
%Quantum Tomography of Semi-Algebraic Sets with Constrained Measurements
\bibitem{Kech15b} M. Kech, Preprint  arXiv:1508.00522 (2015).
%Explicit Frames for Deterministic Phase Retrieval via PhaseLift
\bibitem{Baldwin15} C. H. Baldwin I. H. Deutsch, and A. Kalev, Preprint  arXiv:  1510.02736 (2015).
%Informational completeness in bounded-rank quantum-state tomography
\bibitem{Kueng15} R. Kueng, International Conference on Sampling Theory and Applications (IEEE), pp 402-406 (2015).
%Low rank matrix recovery from few orthonormal basis measurements
\bibitem{Acharya15} A. Acharya, T. Kypraios, and M. Guta, , Preprint  arXiv:  1510.03229 (2015).
%Statistically efficient tomography of low rank states with incomplete measurements
\bibitem{Hradil97}  Z. Hradil,  {\em Phys. Rev. A} {\bf 55}, R1561 (1997).
%Quantum-state estimation.
\bibitem{cvx} Software for disciplined convex programming can found
at http://cvxr.com/. 
\bibitem{teo11}  Y. S. Teo, H. Zhu, B.-G. Englert,    J. \v{R}eh{\'a}\v{c}ek, and Z.
Hradil,  {\em Phys. Rev. Lett.} {\bf 107}, 020404 (2011).
%Quantum-State Reconstruction by Maximizing Likelihood and Entropy. 

\end{thebibliography}
\end{document}